\documentclass[twocolumn,amsmath,amssymb,pra]{revtex4}

\usepackage{graphics}
\usepackage{epsfig}
\usepackage{multirow}

\newcommand{\LiRb}{$^{6}$Li-$^{87}$Rb }
\newcommand{\ket}[1]{| #1 \rangle}
\newcommand{\bra}[1]{\langle #1 |}

\def\oper#1{\hat{\rm #1}}
\def\m{\phantom{}}

\begin{document}

  \bibliographystyle{apsrev}

  \title{Two-channel model of photoassociation in the vicinity of a Feshbach resonance}

       \author{Philipp-Immanuel Schneider %, Yulian V.~Vanne 
               and Alejandro Saenz}

       \affiliation{AG Moderne Optik, Institut f\"ur Physik,
         Humboldt-Universit\"at zu Berlin, Hausvogteiplatz 5-7,
         10117 Berlin, Germany}

       \date{\today}

  \begin{abstract}
    We derive the two-channel (TC) description of the photoassociation (PA) process
    in the presence of a magnetic Feshbach resonance and compare to full coupled
    multi-channel calculations for the scattering of $^{6}$Li-$^{87}$Rb.
    Previously derived results [P. Pellegrini {\it et al.}, Phys. Rev. Lett.
    101, 053201 (2008)] are corrected. The PA process is shown to be fully described 
    by two parameters: the maximal transition rate and the point of vanishing 
    transition rate.
    The TC approximation reproduces excellently the PA transition rates of the full
    multi-channel calculation and reveals, e.g., that the enhancement of the rate at a  
    resonance is directly connected to the position of vanishing rate. 
    For the description of two independent resonances it was found that only three parameters 
    completely characterize the PA process.
  \end{abstract}

    \maketitle

 \section{Introduction}
 \label{sec:intro}

 The phenomenon of a magnetic Feshbach resonance (MFR) is widely used for the manipulation
 of systems of ultracold atoms. One field of interest is the combination of the
 photoassociation (PA) of molecules with MFRs. It has been shown, both theoretically and
 experimentally, that the PA transfer rate can be significantly increased in the vicinity of an
 MFR \cite{ cold:abee98,cold:cour98,cold:gris07,cold:junk08,cold:pell08}. %cold:rega03,cold:deig09,
 This leads to the prospect of creating a large number of ultracold molecules out of a sample 
 of ultracold atoms. These molecules are of great interest for applications in quantum 
 information processing \cite{cold:mich06, cold:rabl06}, the exploration of lattices of 
 dipolar molecules \cite{cold:pupi08}, 
 %, precision measurement of fundamental constants \cite{cold:zele08}, 
 or ultracold chemical reactions \cite{cold:chin05,cold:tsch06}.
%
% Alternatively, Feshbach molecules created by an adiabatic sweep over an MFR have been used especially
% in STIRAP schemes to associate ultracold molecules \cite{cold:danz08,cold:lang08}. However, 
% as pointed out in \cite{cold:kuzn09} Feshbach molecules are usually short-lived such that it could
% be preferable to perform STIRAP instead in the vicinity of an MFR.
% An alternative way to associate ultracold molecules is offered by the photoassociative stimulated 
% Raman adiabatic passage (STIRAP) \cite{cold:danz08,cold:lang08}. This technique usually uses 
% Feshbach molecules created by an adiabatic sweep over an MFR. However, as pointed out in 
% \cite{cold:kuzn09} Feshbach molecules are usually short-lived because of inelastic collisions with
% other atoms and molecules. It is thus proposed to perform STIRAP not with Feshbach molecules but
% in the vicinity of an MFR.

 For all PA schemes that exploit the enhancement of PA at an MFR, the understanding of the 
 interplay between both processes is important. Here, we seek to describe the process by a
 two-channel (TC) approximation \cite{cold:fesh58}. In \cite{cold:pell08} this approximation has 
 recently been used to predict the behavior of the PA transition
 rate as a function of the scattering length. We review this approach and find a simplified expression 
 with only two instead of three free parameters. These two parameters which can be, e.g., the maximal 
 transition rate and the position of the minimal transition rate can be obtained either from multi-channel 
 (MC) calculations or from experimental observations. They can serve as a classification of transition 
 processes and reveal a universal dependence of the enhancement of the transition rate on the position 
 of vanishing transition rate.

 The Hamiltonian of relative motion for two colliding ground-state alkali-metal atoms 
 is given by \cite{cold:moer95}
 \begin{equation}
   \oper{H} = \oper{T}_{\mu}+\sum_{j=1}^2
   (\oper{V}_j^{\rm hf}+\oper{V}_j^{\rm Z})+\oper{V}_{\rm int}
   \label{eq:BfieldHam}
 \end{equation}
 where $\oper{T}_{\mu}$ is the kinetic energy and $\mu$
 is the reduced mass. The hyperfine operator $\oper{V}_j^{\rm hf}=
 \frac{a_{\rm hf}^j}{\hbar^2}\vec s_j\cdot \vec i_j$ and the Zeeman operator 
 $\oper{V}_j^{\rm Z}=(\gamma_e \vec s_j - \gamma_n \vec i_j)
 \cdot \vec B$ in the presence of a magnetic field $\vec B$ 
 depend on the electronic spin $\vec s_j$, the nuclear spin $\vec i_j$,
 the hyperfine constant $a_{\rm hf}^j$ of atom $j=1,2$, and on the 
 nuclear and electronic gyromagnetic factors $\gamma_n$ and $\gamma_e$.
 The central interaction 
\begin{equation}
  \oper{V}_{\rm int}(R) = V_0(R) \oper{P}_0 + V_1(R) \oper{P}_1
  \label{eq:v01}
\end{equation}
 is a combination of singlet and triplet Born-Oppenheimer potentials
 $V_0(R)$ and $V_1(R)$ where $\oper{P}_{0}$ and $\oper{P}_{1}$ project respectively on the
 singlet and triplet components of the scattering wave function.

 For low collision energies the eigenfunctions of Hamiltonian (\ref{eq:BfieldHam}) may be written 
 as a superposition
\begin{equation}
\label{eq:MCWF}
  \ket{\Psi} = \sum_{\alpha}\Psi_\alpha(R) \ket{\alpha}
\end{equation}
 of $s$-wave functions in the atomic basis $\ket{\alpha} = \ket{f_1,m_{f_1}}\ket{f_2,m_{f_2}}$. 
 Here, $\vec{f}_j=\vec s_j + \vec i_j$ is the total spin of atom $j$ and $m_f$ its projection
 onto the $B$-field axis. 

 In the following we consider an elastic collision, where only the entrance channel with spin 
 configuration $\ket{\alpha_0}$ is open and all other coupled channels are closed. 
 %That is, all channel functions $\Psi_{\alpha}(R)$ apart from 
 %the entrance channel function decay exponentially for large $R$. 

 %The TC approximation is very successfully used to describe resonance 
 %phenomena in MC problems \cite{cold:fesh58,cold:pell08}. It is briefly introduced while a 
 %more rigorous introduction may be found in \cite{cold:frie91,cold:marc04}.

 Within the TC approximation of the scattering process one projects the full MC Hilbert space 
 onto two subspaces, the one of the closed channels (with projection operator $\oper Q$) and the 
 one of the open entrance channel (with projection operator $\oper P$) \cite{cold:fesh58}. 
 The resulting TC Schr\"odinger equation reads
 \begin{eqnarray}
 \label{eq:two_channel_1}
  (\oper H_P -E) \ket{\Psi_P} + \oper W \ket{\Psi_Q} &=& 0 \\
 \label{eq:two_channel_2}
  (\oper H_Q -E) \ket{\Psi_Q} + \oper W^\dagger \ket{\Psi_P} &=& 0\,, 
 \end{eqnarray}
 with $\oper H_P = \oper P \oper H \oper P$, $\oper H_Q = \oper Q \oper H \oper Q$,
 $\oper W = \oper P \oper H \oper Q$, $\ket{\Psi_P}=\oper P \ket{\Psi}$ and
 $\ket{\Psi_Q}=\oper Q \ket{\Psi}$.
 
 An MFR occurs, if the energy $E$ of the system is close to the eigenenergy
 $E_0$ of a bound state $\ket{\Phi_b}$ of the closed-channel subspace.
 %By the help of a variation of the magnetic field $B$, $E_0=E_0(B)$ can be 
 %brought to resonance with the total energy $E$.
 Following the solution in \cite{cold:frie91} we assume that the 
 closed-channel wave function is simply a multiple $A$ of the bound state $\ket{\Phi_b}$,
 i.e. $\ket{\Psi_Q}= A\, \ket{\Phi_b}$. This is equivalent to the usual one-pole approximation.
 Equation (\ref{eq:two_channel_1}) may be solved via the Greens operator 
 $\oper G_P = (E +i \epsilon - \oper H_P)^{-1}$ with $\epsilon\rightarrow 0^+$.

 The general solution thus reads 
 \begin{eqnarray}
  \label{eq:sol2Ch_1}
  \ket{\Psi_P} &=& C \ket{\Phi_{\rm reg}} + A \oper G_P \oper W \ket{\Phi_b}\,, \\
  \label{eq:sol2Ch_2}
  \ket{\Psi_Q} &=& A \ket{\Phi_b}
 \end{eqnarray}
 where $C$ is a normalization constant and $\ket{\Phi_{\rm reg}}$ is the regular solution
 of $\oper H_P \ket{\Psi} = E \ket{\Psi}$.

 Following the procedure given in \cite{cold:frie91} one arrives at the
 closed-channel admixture
 \begin{equation}
   \label{eq:A2Ch}
   A = - \tilde{C} \sqrt{\frac{2}{\pi \Gamma}} \sin \delta_{\rm res}\,,
 \end{equation}
 with the normalization constant $ \tilde C = C / \cos \delta_{\rm res}$.
 The Greens operator $\oper G_P$ explicitly given in \cite{cold:frie91} yields 
 a long range behavior of the open channel
 \begin{equation}
 \label{eq:Asymp_OC}
  \left.\Psi_P(R)\right|_{R \rightarrow \infty} =
  \tilde C \sqrt{\frac{2 \mu}{\pi \hbar^2 k}} 
  \sin(k R + \delta_{\rm bg} + \delta_{\rm res})\,.
 \end{equation}
 The total phase shift $\delta = \delta_{\rm bg} + \delta_{\rm res}$ results from the 
 background phase shift $\delta_{\rm bg}$ of the regular solution $\ket{\Phi_{\rm reg}}$, 
 which is connected to the background scattering length $a_{\rm bg} = -\lim_{k\rightarrow 0}
 \tan{\delta_{\rm bg}}/k$,
 and a contribution $\delta_{\rm res}$ due to the resonant coupling to the bound state.  
 The resonant phase shift has the functional form
 \begin{equation}
 \label{eq:delta_res_vs_E}
  \tan \delta_{\rm res} = -\frac{\Gamma/2}{E-E_{\rm R}}
 \end{equation}
 where $E_R=E_0+\bra{\Phi_b} \oper W^\dagger \oper G_P \oper W \ket{\Phi_b}$ 
 lies close to the energy of the bound state $E_0(B)$. 
 The width of the resonance is given by 
 $\Gamma = 2\pi |\bra{\Phi_b} \oper W \ket{\Phi_{\rm reg}}|^2$.

 One assumes that $E_R$ depends approximately linearly
 on the magnetic field, i.e. for $E\rightarrow 0$ there is some $\sigma$ and $B_R$ such 
 that $E_R=\sigma(B-B_R)$. % with $\sigma=\frac{\rm{d} E_R}{\rm{d} B}(B_R)$.
 This yields with $\Delta B\equiv\Gamma (2 k a_{\rm bg} \sigma)^{-1}$ the well known relation
 \cite{cold:moer95a}
 \begin{equation}
 \label{eq:a_vs_B}
  a_{\rm sc} = a_{\rm bg}\left(1 + \frac{\Delta B}{B - B_{R}} \right)
 \end{equation}
 which allows to determine the
 resonant phase shift 
 \begin{equation}
 \label{eq:dres_vs_B}
  \tan \delta_{\rm res} = \frac{k a_{\rm bg} \Delta B}{B_{R}-B}
 \end{equation}
 from experimentally accessible quantities.

 Equipped with the MC solution a convenient way to calculate transition rates to molecular bound states 
 is to transform the scattering wave function of Eq.~(\ref{eq:MCWF}) to the molecular basis 
 $\ket{\chi}=\ket{S,M_S}\ket{m_{i_1},m_{i_2}}$ where $S$ and
 $M_S$ are the quantum numbers of the total electronic spin and its 
 projection along the magnetic field and $m_{i_1}, m_{i_2}$ are the nuclear spin
 projections of the individual atoms. 

 Within the dipole approximation with electronic dipole transition moment $D(R)$ 
 the free-bound transition rate  $\Gamma(B)$ to the final molecular state
  $\ket{\Psi_{\rm f}} = \frac{\Psi_{\nu}(R)}{R}Y_J^M(\Theta,\Phi) \ket{\chi_{\rm f}}$ with 
 vibrational quantum number $\nu$ and rotational quantum number $J$ is then proportional to 
 the squared dipole transition moment
 \cite{cold:sand71}
 \begin{equation}
   \label{eq:Iv}
   I_{\rm MC}(B) = 
            \left|
            \int\limits_0^{\infty}
            \Psi_\nu(R)
            D(R)
            \psi_{\chi_{\rm f}}(R) dR
            \right|^2\,.
 \end{equation}
 Selection rules allow only transitions from the 
 $s$-wave scattering function to a final state with $J=1$.
 Due to the orthogonality of the molecular basis, only one molecular channel 
 $\psi_{\chi_{\rm f}}(R)$ with the same spin state as the final state has to be considered.
  
 The solutions~(\ref{eq:sol2Ch_1},\ref{eq:sol2Ch_2}) of the TC approximation %formulated in the atomic basis 
 yield together with the behavior of the closed channel admixture in (\ref{eq:A2Ch}) 
 a squared dipole transition moment
 \begin{eqnarray}
   \label{eq:Iv2Ch}
   \nonumber
   I_{\rm TC}(B) &=& |\bra{\Psi_{\rm f}}\oper D \ket{\Psi}|^2 \\
   &=& |\tilde C|^2
   \left|
     C_1 \cos \delta_{\rm res} -
     C_2 \sin \delta_{\rm res}
   \right|^2\,,
 \end{eqnarray}
 where 
 $C_2 = \sqrt{2/\pi\Gamma}(\bra{\Psi_{\rm f}}\oper D \oper G_P \oper W \ket{\Phi_b} +
 \bra{\Psi_{\rm f}}\oper D \ket{\Phi_b})$ and 
 $C_1 = \bra{\Psi_{\rm f}}\oper D \ket{\Phi_{\rm reg}}$
 do not vary with the
 magnetic field $B$. %within the TC approximation. 
 The prefactor $\tilde C$
 may vary with $a_{\rm sc}$. However, in the following we consider energy-normalized
 scattering solutions with $\tilde C \equiv 1$.
 Introducing $ \beta =|\tilde C|^2( C_1^2 + C_2^2)$ and $\tan \delta_0 = C_1/C_2$
 one can further simplify Eq.~(\ref{eq:Iv2Ch}) to
 \begin{equation}
   \label{eq:IvSimp2Ch}
   I_{\rm TC}(B) =  \beta
    \sin^2 \left(\delta_{\rm res} - \delta_0\right)\,.
 \end{equation}
 Due to the low-energy behavior of the
 regular solution $\Gamma\propto k$ and $C_1\propto \sqrt{k}$ holds for $k\rightarrow 0$
 \cite{cold:moer95}. Therefore one can associate a finite length 
 \begin{eqnarray}
 a_e = - \lim_{k\rightarrow 0}\frac{\tan \delta_0}{k}
 \label{eq:a_e}
 \end{eqnarray}
 with the phase shift $\delta_0$. 

 Let us point out some differences to the previously derived result for the dipole transition 
 moment in \cite{cold:pell08}. In the notation of the current work Eq.~(8) in \cite{cold:pell08} gives
 \begin{equation}
   \label{eq:IvSimpCote}
   I_{\rm TC}(B) =  K |1 + C_1 \tan\delta_{\rm res} + C_2 \sin\delta_{\rm res}|^2 
 \end{equation}
 with $K=|\bra{\Psi_{\rm f}}\oper D \ket{\Phi_{\rm reg}}|^2$, 
 $C_1 = \bra{\Psi_{\rm f}}\oper D \ket{\Phi_{\rm irr}}/\bra{\Psi_{\rm f}}\oper D \ket{\Phi_{\rm reg}}$,
 and
 $C_2 = -\sqrt{2/\pi\Gamma}\bra{\Psi_{\rm f}}\oper D \ket{\Phi_b}/
 \bra{\Psi_{\rm f}}\oper D \ket{\Phi_{\rm reg}}$.
 The most obvious difference to Eq.~(\ref{eq:Iv2Ch}) is the dependence on three parameters $K,C_1$ 
 and $C_2$ and not just two.
 This is a result of an inconsistent normalization of open and closed channels in \cite{cold:pell08}. 
 The open channel was not energy normalized and leads thus to a term proportional to $\tan\delta_{\rm res}$.
 Furthermore, the open channel was described as a pure sum of regular and irregular solution.
 This may, however, only be done for interatomic distances, were the coupling to 
 the closed channels induced by the exchange energy is negligible. 
% A transition to states within the same
% electronic configuration as considered in \cite{cold:pell08} depends unfortunately on the scattering wave
% function at distances where the exchange energy is considerable.
 A fit of Eq.~(\ref{eq:IvSimpCote}) to a full MC calculation seemed to be nevertheless possible which 
 might however be the result of the freedom of three fitting parameters. 

 The universal dependence of the PA transition rates on just two parameters in Eq.~(\ref{eq:IvSimp2Ch}) 
 reveals an important physical aspect. The enhancement of the transition rate is directly connected to 
 the position of vanishing transition rate.
 On the one hand the transition rate is vanishing at a scattering length
 $a_{\rm sc}^{({\rm min})} = a_{\rm bg} + a_e$ or at a corresponding magnetic field
\begin{equation}
 B_{\rm min} =B_R + \Delta B \frac{a_{\rm bg}}{a_e}\,.
\end{equation}
 On the other hand the point of vanishing transition rate is connected to the enhancement ratio 
\begin{equation}
\label{eq:enhance}
 \frac{\Gamma_{\rm max}}{\Gamma_{\rm bg}} = \frac{1}{\sin^2\delta_0} 
 \approx \frac{1}{k^2 a_e^2} = 
 \left(\frac{B_{\rm min}-B_R}{k\, a_{\rm bg} \Delta B}\right)^2
\end{equation}
 of the maximum transition rate $\Gamma_{\rm max}$ and the background rate $\Gamma_{\rm bg}$ 
 in the presence of an off-resonant magnetic field (i.e. where $\delta_{\rm res}=0$). 

\begin{figure}[ht]
\centering
     \includegraphics[width=0.45\textwidth]{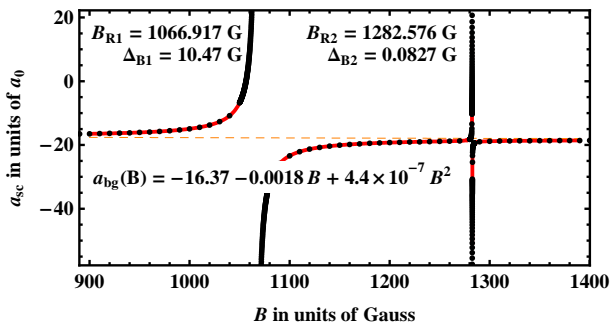}
 \caption{{\footnotesize
     (Color online)
     Scattering length $a_{\rm sc}$ as a function of the external magnetic field 
     value $B$ for \LiRb scattering at $E=50\,$Hz (dots). A fit according to 
     Eq.~(\ref{eq:a_vs_B_2R}) is depicted by the solid (red) line. 
     The value of $a_{\rm bg}(B)$ is shown by the dashed (orange) line.
     All fitting parameters are shown in the plot.
}}
\label{fig:A_vs_B}
\end{figure}

 In order to verify the TC description of the PA process, we consider the exemplary case of an 
 elastic collision of \LiRb ($^6$Li is atom 1, $^{87}$Rb is atom 2)
 in the initial atomic basis state $\ket{\alpha_0}=\ket{1/2,1/2}\ket{1,1}$.
 For an energy 50 Hz above the threshold of the entrance channel which is well in the 
 $s$-wave scattering regime the MC solution was calculated for different magnetic 
 fields $B$ in \cite{cold:gris09a}. For $B<1500\,$G two $s$-wave resonances occur, 
 a broad one at $B=1066,917\,$G which was also observed experimentally \cite{cold:deh08}, and 
 a narrow one at $B=1282.576\,$G. 
 The dependence of the scattering length $a_{\rm sc}$ on the magnetic field strength is shown in 
 Fig.~\ref{fig:A_vs_B}.

 Assuming that the two resonances are sufficiently separated in order 
 to describe the process by two independent resonances one may generalize Eq.~(\ref{eq:a_vs_B}) 
 to
 \begin{equation}
 \label{eq:a_vs_B_2R}
  a_{\rm sc}(B) = a_{\rm bg}\left(1 + 
  \frac{\Delta B_1}{B - B_{R1}} + 
  \frac{\Delta B_2}{B - B_{R2}}\right)\,.
 \end{equation}
 Additionally, we account for effects beyond the one-pole approximation by allowing $a_{\rm bg}$
 to vary slowly with $B$ as $  a_{\rm bg}(B) = a_0 + a_1\cdot B + a_2 \cdot B^2$. 
 With this quadratic expansion, a fit according to Eq.~(\ref{eq:a_vs_B_2R}) excellently reflects the
 MC behavior as shown in Fig.~\ref{fig:A_vs_B}.

\begin{figure}[t]
\centering
     \includegraphics[width=0.46\textwidth]{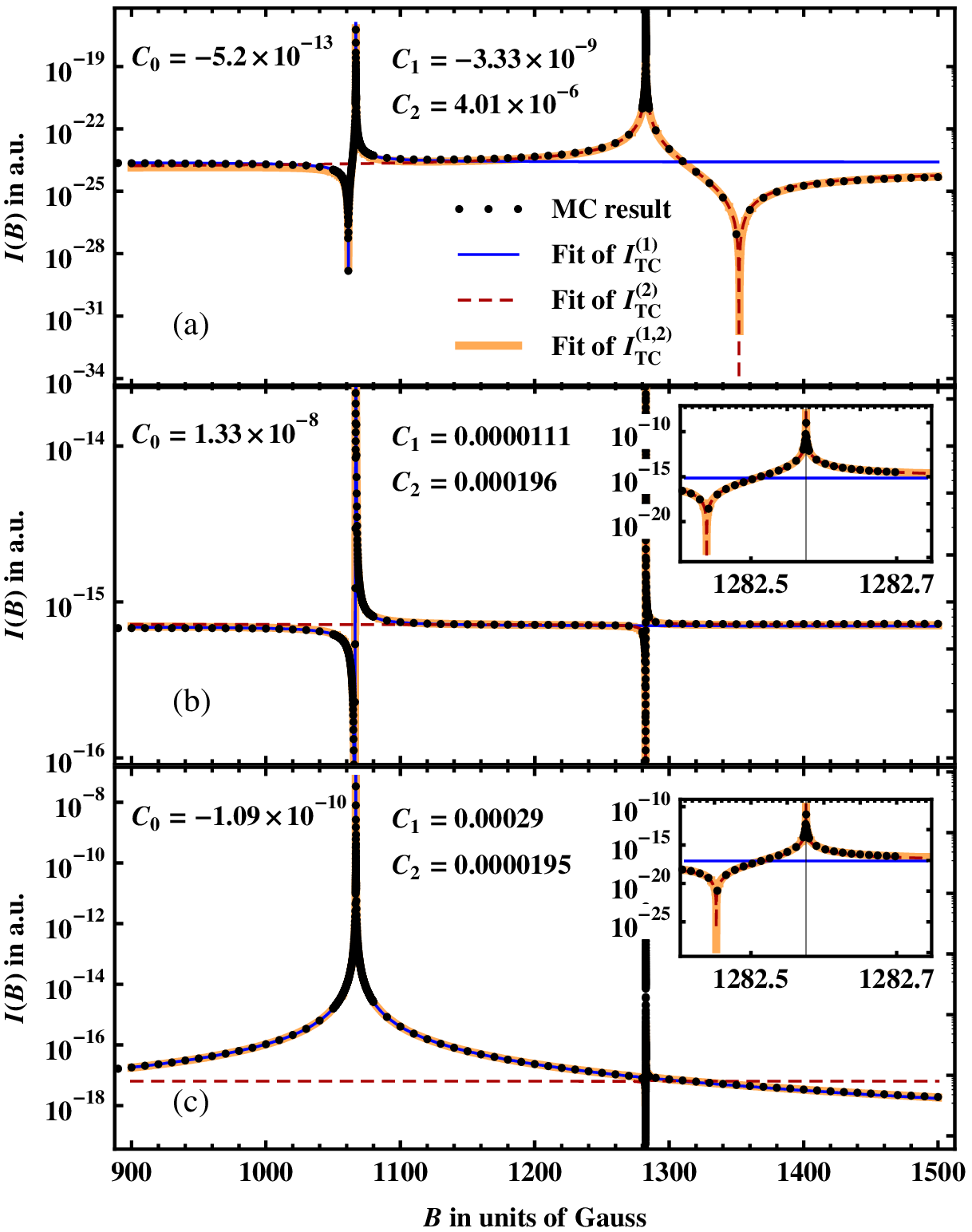}
 \caption{{\footnotesize
     (Color online)
     Squared dipole transition moment as a function of the external magnetic field 
     for \LiRb scattering at $E=50\,$Hz (dots). Transitions to the molecular 
     singlet state $\ket{0,0}\ket{0,3/2}$ (a) and the triplet states
     $\ket{1,-1}\ket{1,3/2}$ (b) and $\ket{1,1}\ket{1,-1/2}$ (c) are considered. 
     In each graph a fit according to Eq.s~(\ref{eq:TC12}) and (\ref{eq:TC1TC2}) is performed and
     the resulting fitting parameters for Eq.~(\ref{eq:TC12}) are shown.
     The insets focus respectively on a region $B_{R2}-2\Delta B_2\leq B\leq B_{R2}+2\Delta B_2$.
}}
\label{fig:I_vs_B}
\end{figure}
 
 We consider the exemplary case of a dipole transitions of the scattering state to the absolute 
 vibrational ground state of the electronic singlet configuration $X^1\Sigma^+$ and the triplet 
 configuration $a^3\Sigma^+$. These transitions take place at internuclear distances where the coupling
 between all atomic channels is strong such that any deficiency of the TC description
 would become obvious.
 The MC rate was calculated in \cite{cold:gris09a} for an electronic dipole moment in the linear 
 approximation $D(R) = D_0 + D_1 \cdot R$, where $D_0$ could be neglected.
 In the following we use $D_0=0$ and $D_1=E_h/a_0$. 
 The magnetic field dependence of the according dipole transition moment to ground states in 
 different spin configurations is shown in Fig.~\ref{fig:I_vs_B}.

 We fit the MC behavior by assuming again that the two resonances are sufficiently separated, 
 such that one can add the transition amplitudes 
 $A_{\rm TC}^{(j)}(B) = C_0 \cos(\delta_{\rm res}^{(j)}) - C_j \sin(\delta_{\rm res}^{(j)})$
 %  A_{\rm TC}^{(2)}(B) = C_1 \cos(\delta_{\rm res}^{(2)}) - C_3 \sin(\delta_{\rm res}^{(2)})
 of both resonances $j=1,2$ independently and take the absolute square of the sum
 \begin{equation}
  \label{eq:TC12}
   I_{\rm TC}^{(1,2)}(B) = |A_{\rm TC}^{(1)}(B) + A_{\rm TC}^{(2)}(B)|^2\,
 \end{equation}
 in order to determine the dipole transition moment. The resonant phase shifts
 $\tan \delta_{\rm res}^{(j)} = k a_{\rm bg}(B) \Delta B_j/(B_{Rj}-B)$
 for $j=1,2$ are in analogy to Eq.~(\ref{eq:dres_vs_B}) associated to the resonant coupling to 
 two different closed-channel bound states. 
 Since $C_0$ does not depend on the resonant molecular bound state $\ket{\Phi_b}$, 
 it is independent of the magnetic field $B$. Hence, for describing a transition process to a specific
 molecular state for two well separated resonances, one needs only three independent parameters.

 %In order to determine in what magnetic field range the TC approximation of a single resonance is capable 
 %of reproducing the MC results, 
 Additionally a fit to the behavior
 \begin{equation}
  \label{eq:TC1TC2}
   I_{\rm TC}^{(j)}(B) = \beta_j \sin^2(\delta_{\rm res}^{(j)}-\delta_0^{(j)})%, \qquad
%   I_{\rm TC}^{(2)}(B) = \beta_2 \sin^2(\delta_{\rm res}^{(2)}-\delta_0^{(2)}),
 \end{equation}
 for $j=1,2$ is performed which neglects respectively one resonance. 
 Again one can connect the phase shifts $\delta_0^{(j)}$ to the corresponding 
 lengths $a_e^{(j)}$ via Eq.~(\ref{eq:a_e}) which
 allows also to validate the applicability of Eq.~(\ref{eq:enhance}) for both 
 resonances separately.

 Considering Fig.~\ref{fig:I_vs_B} one finds that for all shown transitions the TC approximation for two
 well separated resonances excellently describes the magnetic-field dependence of the MC transition rate. 
 The behavior of Eq.~(\ref{eq:TC1TC2}) reproduces each of the MC resonances $j=1,2$ in an interval of 
 several $\Delta B_j$ around the resonant magnetic fields $B_{Rj}$.
 With only one parameter more than for the description of a single resonance, both resonances can be 
 well described by Eq.~(\ref{eq:TC12}). This is remarkable, since the behavior of the transition rates to 
 all four different spin configurations are quite different. Nevertheless all features are reproduced by
 just three parameters.

\begin{table}[t]
  \caption{  \label{tab:VarPar}
  Enhancement of the dipole transition rate between off-resonant and resonant magnetic field
  in MC and TC description for transitions to all eight possible spin states 
  $\ket{S,M_S}\ket{m_{i_1},m_{i_2}}$ at the two resonances at 
  $B_{R1}=1066,917\,$G and $B_{R2}=1282.576\,$G.}
\begin{tabular}{@{}lllll}
\hline 
  Molecular&\multicolumn{2}{c}{Resonance 1} & \multicolumn{2}{c}{Resonance 2}\\
  state &
  $I_{\rm MC}^{\rm max}/I_{\rm MC}^{\rm off}$ & $(k\, a_e^{(1)})^{-2}$ & 
  $I_{\rm MC}^{\rm max}/I_{\rm MC}^{\rm off}$ & $(k\, a_e^{(2)})^{-2}$ \\

\hline
\hline

  $\ket{0, 0}\m\ket{   0,   \frac32}$&
  $5.81\cdot 10^6$   &$5.83\cdot 10^6$         &$6.85\cdot 10^{12}$&$1.39\cdot 10^{13}$\\
  $\ket{0, 0}\m\ket{   1,   \frac12}$&
  $2.62\cdot 10^8$   &$3.49\cdot 10^8$         &$6.57\cdot 10^6$   &$1.27\cdot 10^7$\\
  $\ket{1,-1}\ket{   1,   \frac32}$&
  $1.80\cdot 10^5$   &$1.75\cdot 10^5$         &$5.73\cdot 10^7$   &$5.41\cdot 10^7$\\
  $\ket{1, 0}\m\ket{   0,   \frac32}$&
  $8.86\cdot 10^6$   &$8.69\cdot 10^6$         &$1.63\cdot 10^{13}$&$5.93\cdot 10^{13}$\\
  $\ket{1, 0}\m\ket{   1,   \frac12}$&
  $5.04\cdot 10^{9}$&$2.14\cdot 10^9$          &$2.15\cdot 10^8$   &$4.71\cdot 10^7$\\
  $\ket{1, 1}\m\ket{  -1,   \frac32}$&
  $5.77\cdot 10^6$   &$8.41\cdot 10^6$         &$7.76\cdot 10^{12}$&$5.21\cdot 10^{13}$\\
  $\ket{1, 1}\m\ket{   0,   \frac12}$&
  $1.88\cdot 10^7$    &$2.13\cdot 10^7$        &$1.32\cdot 10^{13}$&$6.57\cdot 10^{13}$\\
  $\ket{1, 1}\m\ket{   1,  -\frac12}$&
  $1.09\cdot 10^{10}$&$1.18\cdot 10^{12}$      &$3.66\cdot 10^7$   &$4.41\cdot 10^7$\\
\hline
      \end{tabular}
\end{table}

 The good description of the MC transition rates by Eq.~(\ref{eq:TC1TC2}) suggests that Eq.~(\ref{eq:enhance}) 
 indeed reflects the dependence of the PA enhancement on the position of vanishing transition rate.
 In Eq.~(\ref{eq:enhance}) the enhancement was defined relative to the transition rate at the background 
 scattering length. %, i.e. where $\delta_{\rm res}=0$. 
 This point is reached, however, only at infinite detuning from the resonant magnetic field $B_R$. 
 In order to nevertheless verify the validity of Eq.~(\ref{eq:enhance}) we relate
 the maximal transition rate $I_{\rm MC}^{\rm max}$ at each resonance separately to the transition rate 
 $I_{\rm MC}^{\rm off}=I_{\rm MC}(800\,{\rm G})$ far away from both resonances. 
 We do not choose a magnetic field with larger detuning to avoid effects of other 
 resonant molecular bound states. %(The MC calculations indicate a broad resonance approaching $B=0$).
 In Tab.~\ref{tab:VarPar} the ratio $I_{\rm MC}^{\rm max}/I_{\rm MC}^{\rm off}$ is compared for both
 resonances to the prediction of Eq.~(\ref{eq:enhance}) for transitions to the vibrational ground states 
 of all eight possible spin configurations in the molecular basis. One finds that the order of magnitude
 generally agrees excellently. Only for few transitions such as the one to the molecular states 
 $\ket{1, 1}\ket{1, -1/2}$ at the first (broad) resonance the orders of magnitude differ significantly. 
 A view on Fig.~\ref{fig:I_vs_B} (c) reveals that this is not related to a break down of
 Eq.~(\ref{eq:enhance}), but that the absence of a vanishing transition rate leads to a comparably 
 slow degradation of the transition rate such that $I_{\rm MC}^{\rm off}$ is not a good representation 
 for the background transition rate. On the other hand, for transitions for which the background 
 transition rate is quickly approached when detuning from the resonance, the two estimates of the
 enhancement agree even to the first significant digit (see the third row in Tab.~\ref{tab:VarPar} and 
 the corresponding Fig.~\ref{fig:I_vs_B} (b)). This and the results above demonstrate that the TC
 approximation provides an excellent basis to understand PA processes in the presence of an external
 magnetic field inducing an MFR.

 We thank Y. V. Vanne for fruitful discussions and for providing us with the differential 
 MC transition rates of \cite{cold:gris09a}. 
 The authors are grateful to the {\it Deutsche Forschungsgemeinschaft} (SFB\,450 C6) for financial support.

% \bibliography{cold,gen,vdw,bsp}

% 
\end{document}